\documentclass[runningheads]{llncs}
\AtBeginDocument{%
  \providecommand\BibTeX{{%
    \normalfont B\kern-0.5em{\scshape i\kern-0.25em b}\kern-0.8em\TeX}}}
\usepackage{siunitx}
\usepackage{amsmath,amsfonts}
\usepackage{algorithm}
\usepackage{xcolor}
\usepackage{moreverb}
\usepackage{tabularx}
\usepackage{threeparttable} %Table
\usepackage{multirow} %Table: multirow
\usepackage{rotating} %Table: rotate text
\usepackage{bigstrut} % Table: extra row height
\usepackage[utf8]{inputenc}
\usepackage{paralist}
\usepackage{lipsum}
\usepackage{soul}
\usepackage[noend]{algpseudocode}
\usepackage{subfig} % for subfigures
\usepackage{caption}
\usepackage{lipsum}

\usepackage[labelsep=period]{caption}
\begin{document}

\title{Detection of Energy Consumption Cyber Attacks on Smart Devices
}
%\author{Anonymous}
%\institute{}

%\begin{comment}

\author{Zainab Alwaisi\inst{1}
\and
Simone Soderi\inst{1,2}
\and
Rocco De Nicola\inst{1,2}
}
\authorrunning{Z. Alwaisi et al.}

\institute{IMT School for Advanced Studies, Lucca, Italy\\
\email{\{zainab.alwaisi, simone.soderi, rocco.denicola\}@imtlucca.it} \and
CINI Cybersecurity Laboratory, Roma, Italy 
}
%\end{comment}

%
\maketitle

\begin{abstract}
With the rapid development of the Internet of Things (IoT) technology, intelligent systems are increasingly finding their way into everyday life and people’s homes. With the spread of these technologies, there is a growing concern about the security of smart home devices. Smart home devices suffer from resource-constrained problems, and these devices and sensors could be connected to unreliable and untrustworthy networks. Nevertheless, securing IoT technology is mandatory due to the relevant data handled by these devices. One of the critical tasks to be solved by the concept of a modern smart home is the problem of preventing energy attacks spread and
the usage of IoT infrastructure. One of the possible approaches to abnormal behavior of IoT devices and IoT cyberattack detection is monitoring energy consumption.
Moreover, building a lightweight algorithm for securing IoT devices is essential to consider the limitation of its resources. This paper presents a lightweight technique for detecting energy consumption attacks on smart home devices based on analyzing the received packets by the smart devices. The proposed algorithm considers three different protocols, TCP, UDP, and MQTT, and different device statuses, like \emph{Idle}, \emph{active}, and when it is under attack. Moreover, it considers the resource constraints of the smart devices for detecting abnormal behaviors and sending an alert to the administrator as soon as the attack is detected. The proposed approach effectively detects energy consumption attacks by measuring the packet reception rate of the smart devices for different protocols.

\end{abstract}

\newpage
\noindent\rule{8.4cm}{1pt}\\
Kindly reference this version of the paper:\\

Alwaisi, Z., Soderi, S., De Nicola, R. (2024). Detection of Energy Consumption Cyber Attacks on Smart Devices. In: Perakovic, D., Knapcikova, L. (eds) Future Access Enablers for Ubiquitous and Intelligent Infrastructures. FABULOUS 2023. Lecture Notes of the Institute for Computer Sciences, Social Informatics and Telecommunications Engineering, vol 542. Springer, Cham. https://doi.org/10.1007/978-3-031-50051-0\_12
\\

You can reference the following BibTeX entry:
\begin{verbatim}
@InProceedings{10.1007/978-3-031-50051-0_12,
author="Alwaisi, Zainab
and Soderi, Simone
and De Nicola, Rocco",
editor="Perakovic, Dragan
and Knapcikova, Lucia",
title="Detection of Energy Consumption Cyber Attacks on Smart Devices",
booktitle="Future Access Enablers for Ubiquitous and Intelligent Infrastructures",
year="2024",
publisher="Springer Nature Switzerland",
address="Cham",
pages="160--176",
isbn="978-3-031-50051-0",
doi="10.1007/978-3-031-50051-0_12"
}






\end{verbatim}
\noindent\rule{8.4cm}{1pt}
\section{Introduction}
The Internet of Things (IoT) can incorporate many heterogeneous devices such as cameras, smart meters~\cite{avula2022}, vehicles, and others transparently while providing open access to various data generated by such devices to provide new services to citizens and companies~\cite{Zanella2014}. The IoT paradigm can be extremely massive and complex. It may contain tens of thousands of sensors, actuators, and gateways. Devices can communicate with gateways via different protocols, whereas gateways may connect with the internet and cloud-based apps via a similarly diverse range of protocols~\cite{rondon2022}. IoT technology's services find applications in many domains such as automotive, medical aids, smart grids, and many others~\cite{dawod2022}. The relevant data exchanged between smart IoT devices are more vulnerable to attacks since they are often deployed in a hostile and insecure environment~\cite{Bellavista2013}. 

In this complex architecture, data can be processed by various heterogeneous entities. Data transmission, security, and integrity are key aspects to be considered. As a result, protocols and technologies are required to provide data security, access management, and flow data transmission~\cite{barani2022}. Many recent studies have been conducted to cope with security issues in the IoT paradigm~\cite{xu2022}~\cite{ang2022}. Some of these studies concentrate on security issues at a particular layer, whereas other approaches aim at providing end-to-end security~\cite{khan2018iot}.
Several methods and protocols have been suggested, primarily concerned with reducing energy consumption and increasing the network lifetime~\cite{birajdar2017}~\cite{rahmadhani2018}. Therefore, security solutions are mandatory to protect IoT devices from intruder attacks. This paper aims to secure low-resource IoT devices, such as smart home devices, against energy consumption attacks~\cite{pattewar2022}.

In smart homes, detecting energy consumption attacks is required to protect the energy from vulnerability threats that could access the home network and attack the smart devices. 
Monitoring the energy consumption of IoT devices is a possible way to detect those performing attacks which require significant energy consumption. In addition, the energy consumption analysis-based approach is more secure when the device's kernel is already compromised. Data integrity cannot be guaranteed once the device has been compromised~\cite{Bobrovnikova2021}.

One of the most critical studies nowadays concerns the efficient use of energy resources. Almost a third of the total energy consumption comprises specific losses; for example, the energy is consumed not on purpose~\cite{USA}. Additional growth in energy consumption is also expected. Increasing awareness of the problems of energy saving and energy efficiency also helps to develop the concept of a modern smart home. Furthermore, at first, this concept was to connect sensors and devices over a network for remote access, monitoring, and control of the living environment and provide the required services to users. While at the present stage, it also involves the optimal use of energy in buildings and malware and IoT cyberattack detection in smart home infrastructure. IoT devices' energy consumption monitoring is a possible way to detect those performing attacks which require significant energy consumption~\cite{rahmadhani2018}, for example, Distributed Denial of Service (DDoS)~\cite{tushir2021} and crypto-mining. 

In this paper, we build a lightweight algorithm that considers the resource constraints for smart devices to detect energy consumption attacks. The algorithm is used to monitor the packet reception rate of the smart devices on different protocols. In this algorithm, we used the following protocols: Transmission Control Protocol (TCP), User Datagram Protocol (UDP), and Message Queue Telemetry Transport protocol (MQTT), as they are popular protocols used nowadays with IoT systems~\cite{Kraijak15,islam2022}. We also consider different devices' statuses, such as \emph{Idle}, \emph{active}, and when they could be under attack. The algorithm automatically fetches the packets' reception rate and divides them into different behaviors, such as normal and abnormal, depending on the presence and absence of the energy consumption attacks. At the same time, the energy consumption of the smart devices is measured to determine the packet reception rate's behavior and to specify whether the packet reception rate's behavior is normal or abnormal. This algorithm successfully detected energy consumption attacks in smart home devices with a cost-efficient experimental setup.

\begin{figure*}[ht]
	\includegraphics[width=\textwidth]{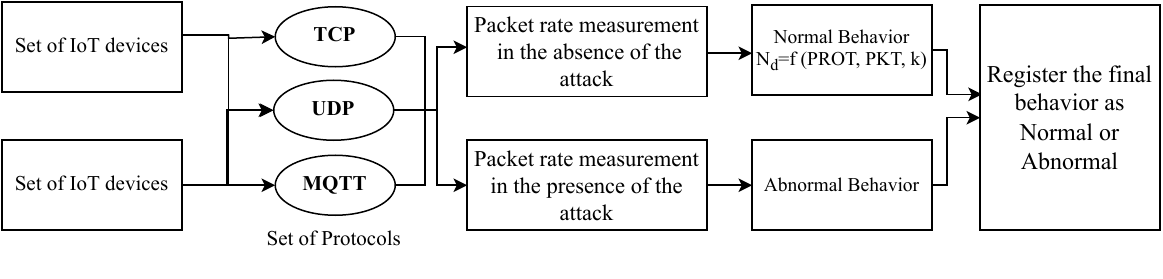}
	\centering
	\caption{Packet Reception Rate measurement in the absence and presence of the attack.}
	\label{FIG:AttackSC1}
\end{figure*}
\subsection{Motivation and Contribution}    
\label{SUBSEC:MOTIVATION}

Security is the main issue that restricts the adoption of IoT in social life. Many researchers have been working to make the IoT a more reliable and secure technology so that it can be adopted in society to make some aspects of human life more manageable and convenient. Since researchers develop many schemes and methods, but due to the constrained environment, e.g., low computational power and low energy of IoT, these techniques are not feasible. Therefore, an added line of protection that considers resource constraints should be built into IoT devices and networks to defend IoT-based organizations from cyber threats. Our main contribution is building a lightweight algorithm to detect energy consumption attacks in smart homes deployed directly at sensors. It applies real-time packet rate measurement to discriminate between smart devices' normal and abnormal packet reception rate behaviors. In this work, we consider three different protocols such as TCP, UDP, and MQTT. We also consider the different device statuses, such as \emph{Idle}, \emph{active}, and when it is under attack, to evaluate the best detection of energy consumption attack. We simulate the detection algorithm and assess the results by applying the proposed algorithm to the smart devices themselves, such as the Raspberry Pi\footnote{https://www.raspberrypi.com/documentation/}. We measure the current consumption of the smart device to monitor the energy while measuring the packet reception rate to discriminate between normal and abnormal behaviors. Therefore, this algorithm design is a protection strategy for IoT devices to maintain their integrity, seamlessly make them available to legitimate users, and protect them from energy consumption attacks by considering their resource constraints.

\subsection{Organization of the paper}    
\label{SUBSEC:OUTLINE}
We organized our paper as follows. Section~\ref{SEC:RW} presents a related work and background reading of energy consumption attacks in IoT systems. We describe our proposal, including metrics definition, methodology, and the detection algorithm, in Section~\ref{SEC:PA}. In Section~\ref{SEC:EXP}, we show the results and discussions. Finally, Section~\ref{ConFU} presents some concluding remarks and future works.

\section{Related Work and Background}
\label{SEC:RW}

Energy-based attacks are often categorised as IoT sensing domain attacks, where the smart devices and sensors are the target~\cite{dabbagh2019}. Dabbagh and Rayes in~\cite{dabbagh2019} described the sensing domain attacks like vampire attacks, jamming attacks, sinkhole attacks, and selective-forwarding attacks. The vampire attack, among others is considered an energy-based attack because it aims to destroy the battery of sensors. The researchers also identified four types of vampire attacks based on the technique used to destroy power: Denial of Sleep, stretch attack, flooding attack, and carousel attack. Patil and Sharma in~\cite{patil2016} also described several Denial of Service (DoS) attacks for wireless sensors. The authors mentioned two attacks that waste the energy of sensors, among others: Denial of Sleep and vampire attacks. Another category of attacks is related to DoS, but they can waste energy indirectly. These are jamming attacks, wormhole attacks, and path-based DoS attacks.

Different authors present detection techniques against energy consumption attacks.
In~\cite{Valentina2017}, it is reported that the primary principle of the energy efficiency approach has been to encourage the use of more efficient smart devices. Home automation control plays a crucial role in efficient and sustainable operation by reducing and identifying energy losses and using energy only when and where it is needed; or by exercising effective control over the operational level of the system for correct application in the proper place. This study~\cite{FORD2017} evaluated well-known home energy management systems in order to identify key differences in functionality and quality by identifying possibilities for energy savings (both behavioral and operational)~\cite{kumar2021secure}. It is also observed that potential benefits related to comfort, convenience, or security can often determine the implementation of energy-efficiency scenarios.
This work~\cite{Shi2019} proposed a detection framework for IoT systems based on energy consumption analysis. The suggested methodology analyzes the energy consumption of smart devices and classifies the monitored devices' attack status, e.g., cyberattacks and physical attacks. A two-stage approach is suggested, with a short time window for rough attack detection and a long time window for fine attack detection.

The authors in~\cite{Felius2020} introduced different techniques to control smart home systems to reduce energy consumption. Feed-forward control is the first approach. By monitoring interference factors in real-time to implement appropriate monitors based on known parameters, such a system directly compensates for interference factors like external wind, solar radiation, and internal heat gain. Another approach is Model-based Predictive Control (MPC)~\cite{Felius2020}, which is used to predict the system's behavior in the future based on models and adjust the system accordingly. Fuzzy logic control does not require a complicated mathematical model to control the system and can be based directly on the quality of user experience. 
In this paper~\cite{Hoffmann2013}, energy consumption analysis approaches were considered, concluding that these approaches do not apply to such devices as smartphones because the typical energy consumption of such devices differs quite a lot in practice. In addition, the noise present in the system by the unpredictable user and environment interactions will lead to many false alarms. 
Also, practical tests were performed, showing that the additional power consumed by malicious applications is too small to be noticeable with the mean error rates of state-of-the-art measurement tools. However, it was noted that DDoS attacks could be detected by studying the energy consumption of similar devices.

The author of this paper~\cite{Bobrovnikova2021} proposes a method for IoT attack detection based on analyzing the smart device's energy consumption, which considers the energy consumption-related user's preference modes. Moreover, it aims to enhance the accuracy of IoT cyberattack detection and localize the IoT malware on these smart devices. The IoT software opcodes sequences study is applied. The proposed technique allows the detection of the performance of the IoT devices, such as DoS and DDoS attacks.

To the best of our knowledge, our work is the first to detect energy consumption attacks in smart home devices, depending on measuring the packet reception rates by the smart devices.

\section{Proposed Algorithm}
\label{SEC:PA}
\begin{figure*}[!ht]
	\includegraphics[width=\textwidth, height=4cm]{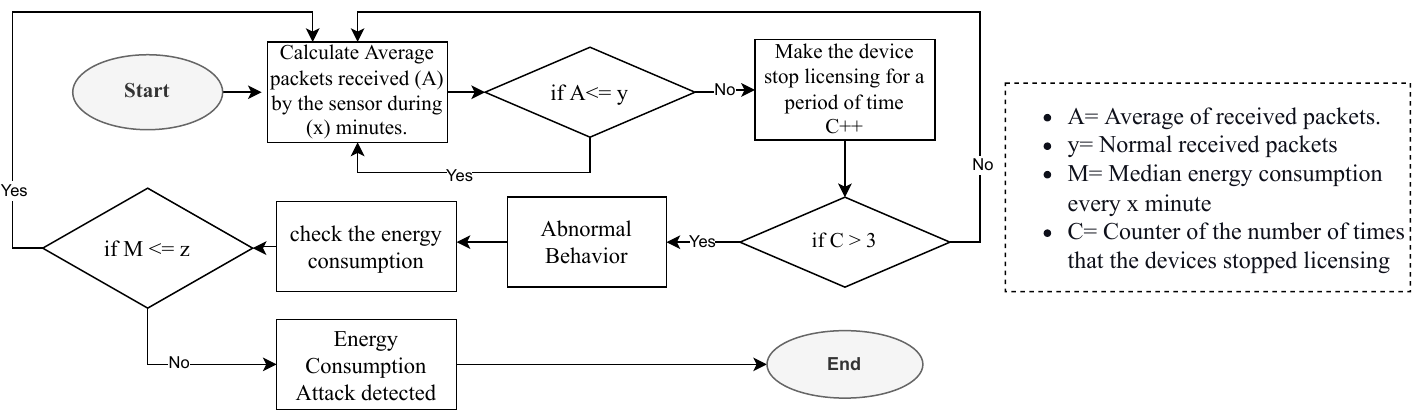}
	\centering
	\caption{A Technique to detect Energy Consumption Attack.}
	\label{FIG:F2D}
\end{figure*}

In this section, we present the algorithm to detect energy consumption attacks in smart home devices by monitoring the packet rate received by the smart devices. The algorithm considers different protocols like TCP, UDP, and MQTT and different device statuses such as \emph{active}, \emph{Idle}, and \emph{under attack}. The proposed algorithm is depicted in Figure~\ref{FIG:F2D}. In our previous work~\cite{BICT2023}, we studied the effect of DDoS, energy consumption DDoS, and Fake Access Points (F-APs) attacks on the energy consumption of the smart healthcare devices.

The algorithm has three phases, 
\begin{inparaenum}[1)]
    \item \emph{collecting phase} where the algorithm collects samples of the number of received packets for different statuses when the device is Idle, active, or under attack; and divides the collected packets from different protocols, e.g., TCP, UDP, and MQTT, into normal or abnormal behaviors;
    \item \emph{calculating phase}, which calculates the collected samples and compares the final results of the fetching packets with the energy measurements to determine whether the state of packets measurements is caused by an energy consumption attack, then divides the final results into (normal, or abnormal behavior); and
    \item \emph{detection phase} where the algorithm applies different conditions to classify if there is an energy consumption attack or not. 
\end{inparaenum}
We build the algorithm inside the Python scripts for automatically fetching the packets and analyzing the normal and abnormal behaviors.

In the detection stage of the proposed technique, the packet reception rate of IoT devices for different protocols is measured and analyzed. If the IoT device has abnormally high received packets, it may have carried out an energy consumption attack. Therefore, smart devices should stop listening to the received packets of such a port. Simultaneously, there should be a counter ($x$) on the total time that the smart device stops listening; if it exceeds ($x$) times, then the algorithm should register it as abnormal behavior. Our algorithm considers the (${x=3}$) times.
\begin{algorithm}
\caption{A Technique to detect Energy Consumption Attack}
\begin{small}
\begin{algorithmic}[1]

\State $N(d)=f(PROT,PKT,k)$
\Comment{\textit{Normal packet reception rate}}
    \State $A=\overline{PKT}$ \Comment{\textit{Received packets in $x$ minutes}}
    \State $y=N(d)$ \Comment{\textit{Normal received packets of smart device $d$}} 
    
    \State Input: $PROT,d$
    \State Output1: Normal($PKT,k,PROT,d$) 
    \State Output2: Abnormal ($PKT,k,PROT,d$) 
    \State Final Result: Output1 $or$ Output2
    \If{$A <= y$} 
        \State return to monitor packets rate
    \Else
        \State Make the device stop licensing for x time
        \State $counter=counter+1$
        \If{$counter>3$}
            \State register the device as abnormal behavior
            \State check energy consumption
        \Else
            \State return to monitor packets rate
        \EndIf
    \EndIf
    
\end{algorithmic}
\end{small}
\end{algorithm}
\subsection{Packet Measurements}
To effectively build a technique to detect energy consumption attacks in IoT systems, it is necessary to take into account the different protocols used for different IoT devices. This algorithm considers three different protocols, e.g., TCP, UDP, and MQTT. Also, it considers different device statuses, e.g., \emph{Idle}, \emph{active}, and when it is under attack.
With the aim of IoT energy consumption attack detection at the learning stage, the packet reception rate of each IoT device in the IoT network in the absence or the presence of an IoT energy consumption attack is measured at a specific interval and at equal sub-intervals of time. Based on these measurements, the number of normal $N(d)$ received packets of IoT devices are constructed, part of them labeled as \emph{normal behavior}  and entered into the database (DB) to deal with them later on.

Let us describe the normal ($N$) packet reception rate measurement in the absence of energy consumption attacks.

\begin{equation} \label{eq2}
N(d) = f(PROT, PKT, k)   \quad \textrm{and} \quad k \in [0,1]
\end{equation}

Where $N(d)$ the normalized receiving packets of an IoT device ($d$) where $d$ the certain smart home device, ($PKT$) received packets at a point in time in the absence of energy consumption attack for a specific protocol ($PROT$), and $K$ is the number of packets measurement in time interval, $n(d)=f(PROT,PKT,k)\in[0,1]$
where $0$ is the minimum received packets, and $1$ is the maximum received packets by the smart devices for a specific protocol.

\subsection{Energy Measurements}
\label{SEC:EM}
With the aim of IoT energy consumption attack detection at the learning stage, the energy consumption measurements of each IoT device in the IoT network in the presence or the absence of IoT energy consumption attacks are measured at a specific interval and at equal sub-intervals of time. Based on these measurements, the number of received packets of IoT devices on $N(d)$ are constructed, part of them labeled as \emph{normal behavior} and others as \emph{abnormal behavior} and entered into the DB to deal with them later on. The energy consumption measurement is essential at the first stage as it is used to determine the behavior of 
the packet reception rate as normal or abnormal.

\begin{figure}[ht]
	\includegraphics[width=\textwidth]{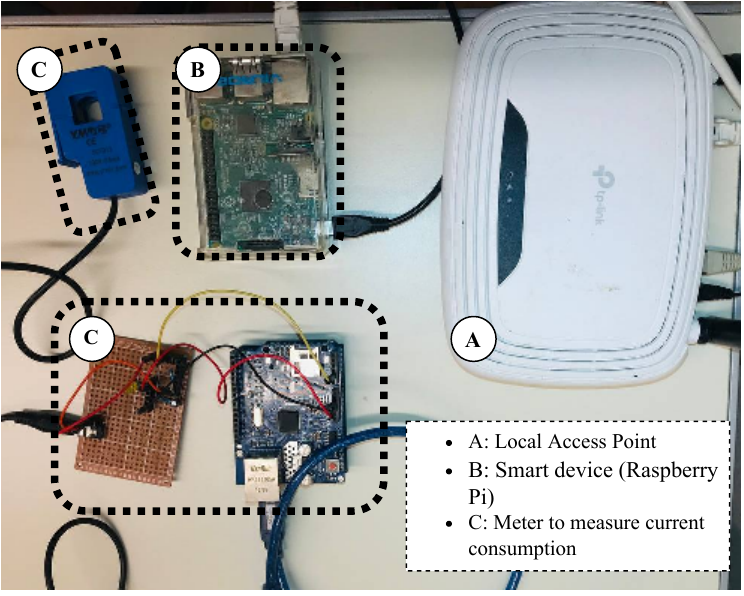}
	\centering
	\caption{Testbed scenario showing the devices used in our experiment and the sensor used to measure the energy consumption.}
	\label{FIG:TCbed}
\end{figure}

With this aim, these IoT devices were infected with malicious attacks, which were able to carry out these types of IoT energy consumption attacks. After that, the energy consumption measurement of each IoT device for different statuses in the presence or the absence of IoT cyberattacks is measured at a specific interval and at equal sub-intervals of time. In this experiment, we designed a smart circuit using a non-invasive current sensor \footnote{https://tinyurl.com/mrxyvr46} with Arduino, capacitors, and other resistors to measure the current consumption of smart home devices. This smart circuit samples voltage, ampere, watt, and current per second. In our experiment, we use the Joule (J) values to calculate the energy consumption of smart devices, as shown in Figure~\ref{FIG:TCbed}. 

Let us describe the energy ($E$) measurement footprints considering the set of different device statuses in the absence or the presence of the attack.
\begin{equation} \label{eq4}
 E(d) = f(e(d),PROT,k) \quad \textrm{and} \quad k \in [0,1]
\end{equation} 
Where ($e_{d}$) the energy measurement ($e$) of the smart device ($d$) at a point in time in
the absence or presence of cyberattacks for a specific protocol ($PROT$), and $K$ is the number of energy measurements in a time interval, $f(e(d),k)\in[0,1]$ where $0$ is the minimum energy consumption measurement, and $1$ presents the maximum energy consumption measurement in the absence or presence of the attack.
\begin{figure}[ht]
	\includegraphics[width=\textwidth]{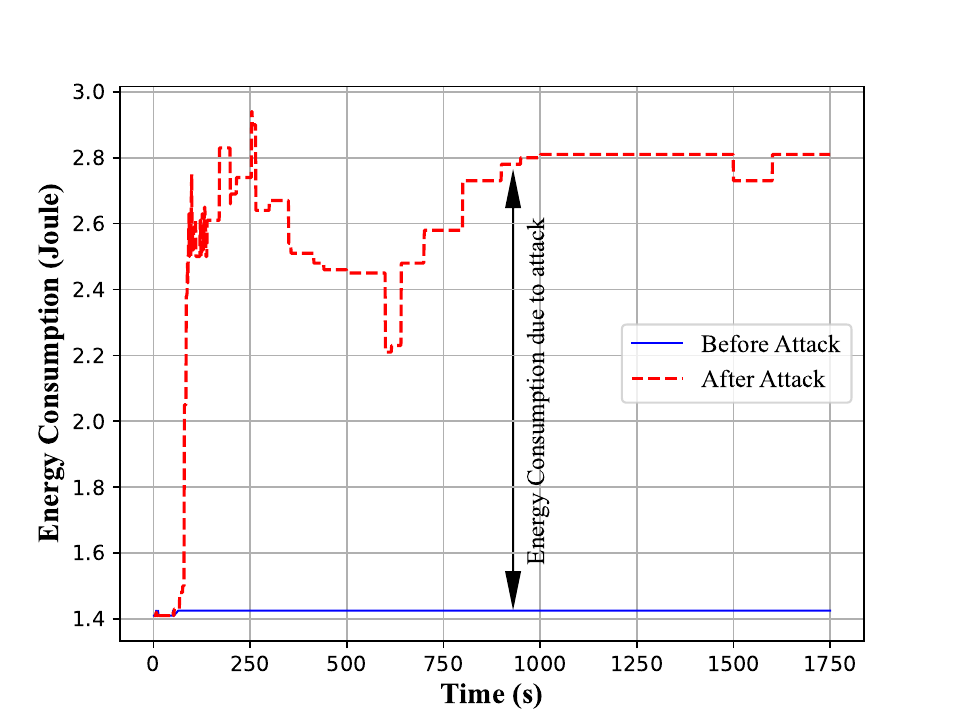}
	\centering
	\caption{Energy consumption measurement of normal and abnormal behaviors of the Raspberry Pi device.}
	\label{FIG:ECRP}
\end{figure}
Therefore, we calculated the energy consumption for every $3$ minutes for a specific smart device; the time for each energy consumption measurement is also registered and entered into the DB.

\subsection{Calculation of normal and abnormal behaviors}

In order to calculate the packet reception rate for each IoT device in normal and abnormal cases, we have divided the code into different parts:\begin{inparaenum}[1)]
\item The first part is to fetch the packet reception rate for each protocol separately, depending on the set of protocols used in our system,
\item We measure the packet reception as shown in Equation~\ref{eq2} for the active smart devices with the absence of the attack and for each protocol separately and register the final results as normal behaviors,
\item Then, we measure the current consumption of the smart device in the case of normal behavior as shown in Equation~\ref{eq4} and monitor the packet reception rate with the energy consumption when the status of the smart device is \emph{On} with the absence of the attack. The monitoring mode continuously fetches the packet, calculates energy for about $30$ minutes, and stores the final results for every $3$ minutes in the DB, 
\item For calculating abnormal behaviors, we send malicious attacks to consume energy for about $30$ minutes to the active smart devices. At this time, we start calculating the energy consumption and the packet reception rate for each protocol separately. Then, we compare the final results with the normal behaviors of such a device. In case of abnormal behavior, we store the final result for every $3$ minutes in the DB as abnormal behaviors,
\item For printing the final results and displaying the normal with the abnormal behaviors, we fetch the stored data from the DB and start calculating the normal with abnormal behaviors,
\item In case there is abnormal behavior with fetching the packets compared to normal behaviors, we notify the system administrator to register the entire case as abnormal behavior.
\end{inparaenum} 

\section{Experimentation and Discussion}
\label{SEC:EXP}
In this section, we describe the testbed scenario that we used to test the algorithm. Also, we show the final results of detecting energy consumption attacks for different protocols using packet reception rate measurement.

\begin{figure}[ht]
	\includegraphics[width=\textwidth]{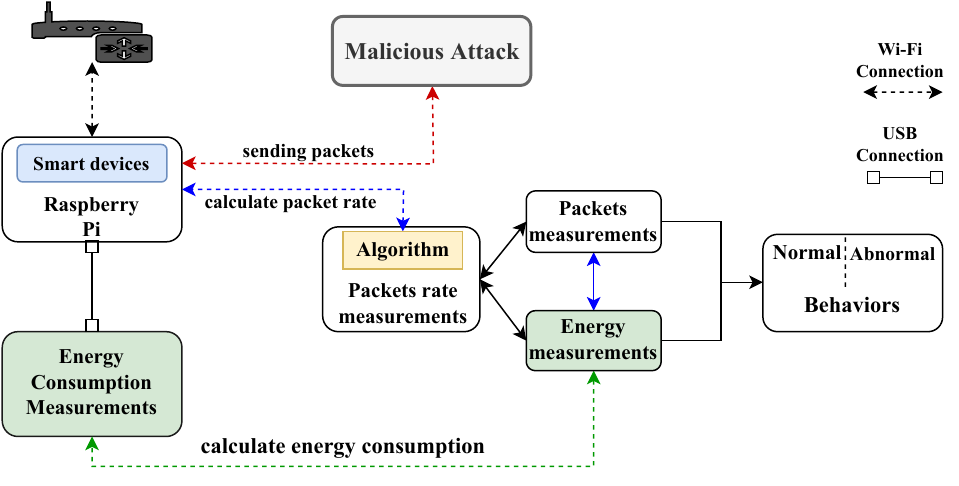}
	\centering
	\caption{Testing Environment.}
	\label{FIG:TCenv}
\end{figure}

\subsection{The testbed scenario}
\label{SEC:Test}

We used Raspberry Pi as a smart home device in this experiment. We used different software tools for attacking data generation and collection. On the adversary side, we used \emph{Nmap}\footnote{https://nmap.org/} to launch a network scan and identify devices’ status, such as \emph{online} or \emph{offline}, IP address, and MAC address. Different tools are used to generate malicious attacks on the victim side, such as \emph{hping3}\footnote{https://www.kali.org/tools/hping3/}. %open-source polymorphic malware creation tool
We designed a smart circuit using a non-invasive current sensor with Arduino, capacitors, and other resistors to measure the current consumption of smart home devices. This smart circuit samples voltage, ampere, watt, and current per second. In our experiment, we use the Joule (J) values to calculate the energy consumption of smart devices. 

We analyze the packet rate received by the smart home device to detect energy consumption attacks. We built a program using \emph{pyshark}\footnote{https://pypi.org/project/pyshark/} to sniff and fetch packets automatically and store the final results in the DB\footnote{https://github.com/developerZA/ATechniuqeToDetectEnergy.git}.

\begin{table}[ht]
    \centering
    \small
    \renewcommand{\arraystretch}{1.6}
    \setlength{\leftmargini}{10pt}
    \begin{footnotesize}
	\caption{Packets analysis depends on protocol type and energy consumption.} 
	\label{tab:PT}
\begin{tabular}{|l|ll|ll|}
\hline
\multicolumn{1}{|c|}{\multirow{2}{*}{\textbf{PROT} }} & \multicolumn{2}{|c|}{\textbf{Normal Behavior}}          & \multicolumn{2}{|c|}{\textbf{Abnormal Behavior}}         \\ \cline{2-5} 
\multicolumn{1}{|c|}{}                      & \multicolumn{1}{|c|}{\text{{Packet}}} & \text{E [J]} & \multicolumn{1}{|c|}{\text{Packet}} & \text{E [J]} \\ \hline
TCP                                         & \multicolumn{1}{l|}{$2000\div6000$}            & $\leq1.42$      & \multicolumn{1}{l|}{$ > 6000$}          & $> 1.42$     \\ \hline
UDP                                         & \multicolumn{1}{l|}{$2000\div6000$}            & $\leq1.42$      & \multicolumn{1}{l|}{$ > 6000$}          & $> 1.42$      \\ \hline
MQTT                                        & \multicolumn{1}{l|}{$2000\div6000$}            & $\leq1.42$      & \multicolumn{1}{l|}{$ > 6000$}          & $> 1.42$        \\ \hline
\end{tabular}
\end{footnotesize}
\end{table}

The total average received packets by the smart devices is calculated by estimating the average rate of the received packets in $30$ minutes compared to the abnormal behavior. We divided the packet reception rate into different slots. For every $3$ minutes, we calculated the average of the received packets in the absence of the attack and stored the final results in the DB as normal behavior. The same calculation is applied to the smart home device when it is under attack. Then the final results are stored in the DB for further calculations. The detection system keeps monitoring the received packets, and in case there are abnormal behaviors received by such a device, we register that case as abnormal behavior.

\begin{figure}[ht]
	\includegraphics[width=\textwidth]{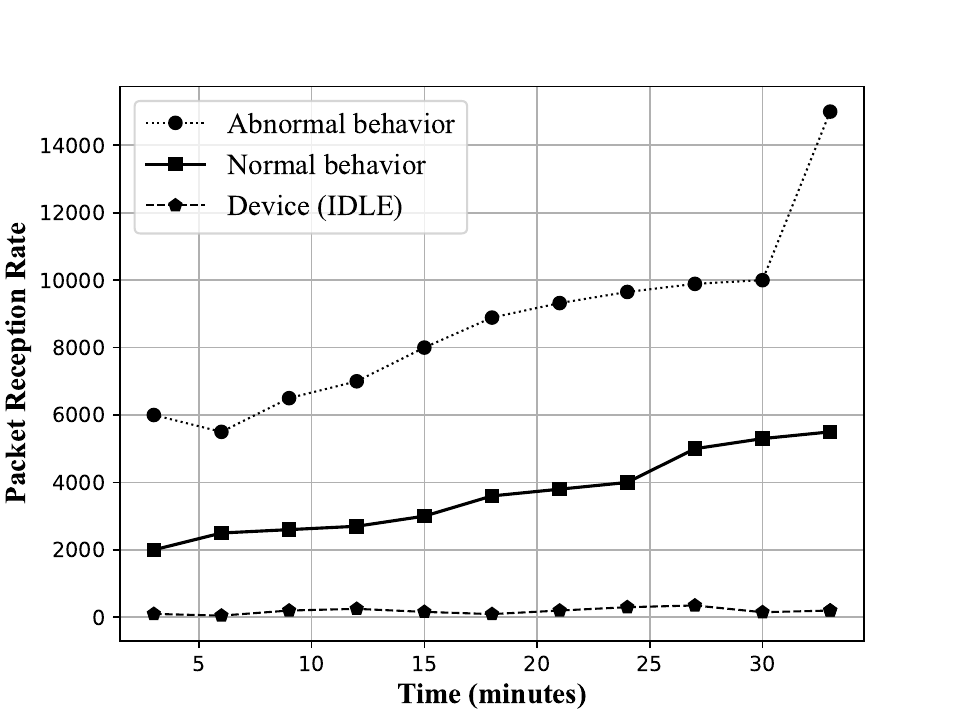}
	\centering
	\caption{Packet reception rate of normal and abnormal behaviors of the TCP protocol.}
	\label{FIG:TCPF}
\end{figure}

To calculate the average received packets by the smart devices of the TCP protocol. We analyzed all the received packets and divided them into different types, such as packets received, re-transmission, and acknowledged. In our experiment, we need the average of the received packet by the smart devices that cause an increase in energy consumption. Then we used the final calculation to detect energy consumption attacks.
%\begin{figure}[ht]
%\centering
%\subfloat[TCP Protocol.]{\label{FIG:TCPF}{\includegraphics[width=0.5\textwidth]{NEW(Faboulous2023)/img/tcp.pdf}}}\hfill
%\subfloat[MQTT Protocol(subscribed packets).]{\label{FIG:MQTT}{\includegraphics[width=0.5\textwidth]{NEW(Faboulous2023)/img/mqtt.pdf}}}\hfill
%\subfloat[UDP Protocol.]{\label{FIG:UDPF}{\includegraphics[width=0.5\textwidth]{NEW(Faboulous2023)/img/udp.pdf}}}
%\caption{Packet reception rate of normal and abnormal behaviors.}
%\label{fig:subfigures}
%\end{figure}
Through the $30$ minutes in the absence or the presence of the attack, we study the received packets by the smart devices for different protocols such as TCP, UDP, and MQTT. Also, we study the total number of times the smart devices stopped listening to understand if energy consumption attacks source the received packets. The normal average of the received packets for TCP protocol in $30$ minutes fluctuates between $2$~k and $5$~k packets, as shown in Figure~\ref{FIG:TCPF}.
%Figure~\ref{FIG:TCPF} also shows the abnormal behavior.

In this experiment, for every $3$ minutes, we calculated the normal and abnormal behavior. So, for the first $3$ minutes, the normal behavior of the received packet is less than $5$~k packets, while the abnormal received packets in the first $3$ minutes in the presence of the attack are more than $6$~k packets. The detection system registers the first case of the first $3$ minutes as abnormal behavior. 
We also calculate the normal and abnormal behaviors for the total of $30$ minutes by calculating the average of the packet reception rates of the normal behaviors and comparing it with the average of the abnormal behaviors to register the entire case of the $30$ minutes as normal or abnormal behavior.

\begin{figure}[ht]
	\includegraphics[width=\textwidth]{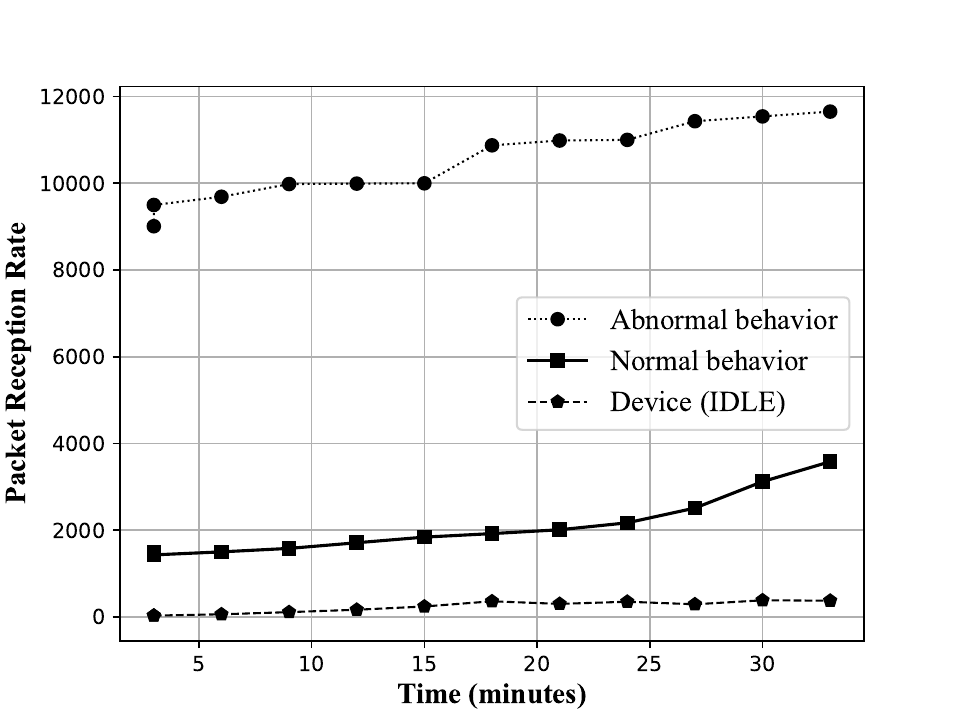}
	\centering
	\caption{Packet reception rate of normal and abnormal behaviors of the UDP protocol.}
	\label{FIG:UDPF}
\end{figure}

\begin{figure}[!ht]
	\includegraphics[width=\textwidth]{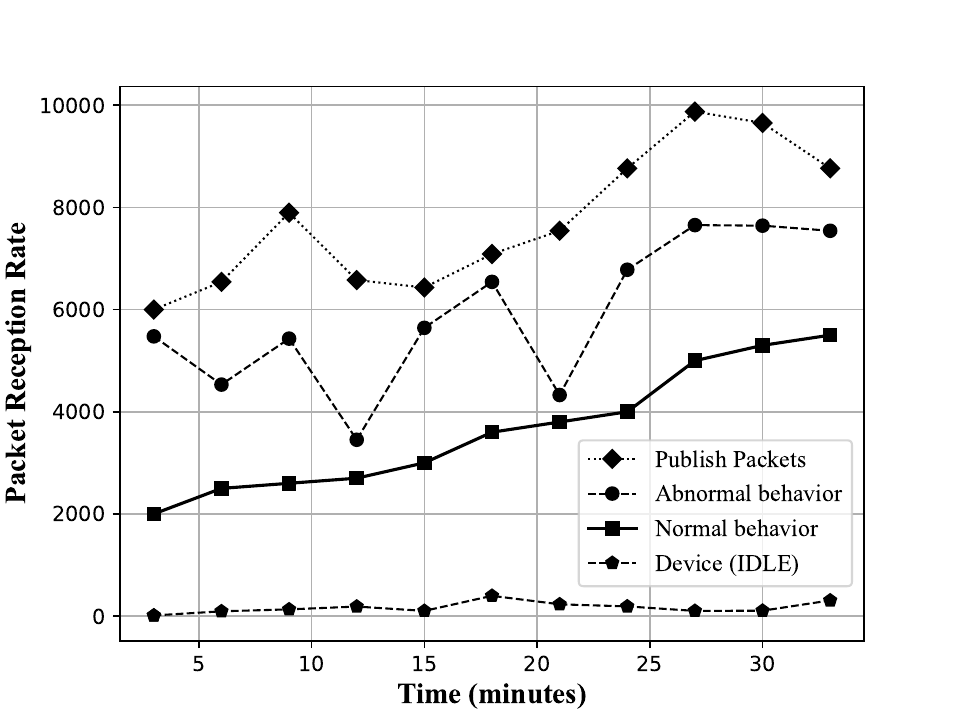}
	\centering
	\caption{Packet subscribed rate of normal and abnormal behaviors of the MQTT protocol.}
	\label{FIG:MQTT}
\end{figure}

\begin{figure}[ht]
	\includegraphics[width=\textwidth]{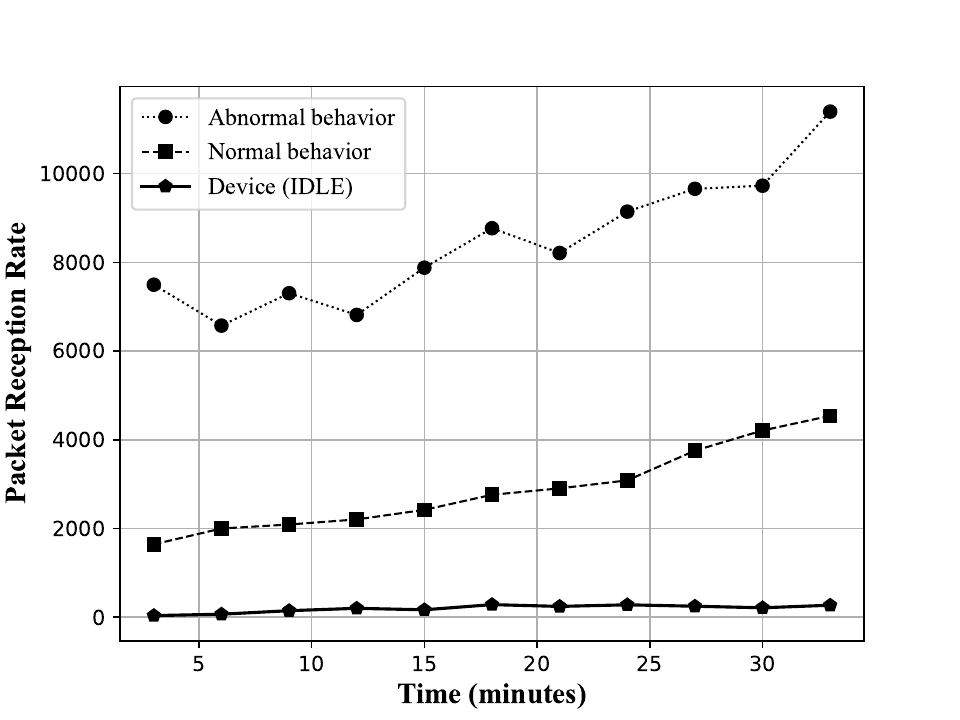}
	\centering
	\caption{General cases (Normal behavior Vs. Abnormal behavior) for TCP, UDP, and MQTT altogether, where the effect of each protocol in normal behavior is as follows: TCP effect is about $45$~\%, and UDP effect about $30$~\%, and the MQTT effect is about $20$~\%. While the impact of TCP is about $40$~\%, MQTT is about $40$~\%, and $20$~\% of UDP is in the presence of the attack.}
	\label{FIG:GNA}
\end{figure}

In the case of UDP protocol, it is not easy to customize the actual receiving packet as the state of such a port cannot be confirmed by network scan using Nmap\footnote{https://nmap.org/} because the port does not send any response. So, in our calculation, as shown in Figure~\ref{FIG:UDPF}, we calculate the normally received packets of UDP protocol by the smart device. We monitor the packet reception rate of the smart devices for $30$ minutes to check the normal and abnormal receiving packets of the Raspberry Pi. The normal behavior of the receiving packets is between $1$~k and $3$~k packets. In contrast, the abnormal behavior of the received packets by the smart device is between $9$~k and more than $12$~k packets.  
Figure~\ref{FIG:MQTT} shows the behavior of the subscribed packet's rate of the MQTT protocol. We study different behaviors of this protocol by registering the number of published and subscribed packets of the smart home device. To detect an energy consumption attack in the case of the MQTT protocol, we consider the number of subscribed packets as they affect the energy resources of the smart devices. 
Therefore, the normal behaviors of the MQTT protocol are registered to be less than $6$~k packets, while the abnormal behaviors reached more than $8$~k packets.
In this algorithm, we also consider the case where we do not have to specify the protocol; by calculating the average received packets for all the used protocols. We find that the normal behavior of the packet reception rate of the Raspberry Pi is between $1500$ packets and less than or equal to $6$~k packets. The abnormal behavior of the total received packets is between $7$~k and more than $12$~k packets, as shown in Figure~\ref{FIG:GNA}.  

\subsection{Experimental Results}
\label{RES}

The IoT device used in this experiment was infected with malicious software and used to carry out different flooding attacks on a target on an isolated network. During the experiments, the energy consumption footprints and the packet reception rate measurements of these IoT devices were obtained under normal operating conditions, as well
as when these smart devices carry out cyberattacks. Each energy consumption footprint and packet reception rate were obtained by taking measurements after $5$~s within $3$ minutes when the smart device performs an attack and normal operation. A total of $30$ minutes of calculation measurement of received packets and the energy consumption footprints of both in the presence of attacks and normal functioning smart devices were built. 

The results of this experiment showed high efficiency of energy consumption attack detection based on the
packet reception rate analysis. At the same time, the
analysis of the packet reception rate for different protocols was considered. As it can be seen from Figure~\ref{FIG:GNA}, the abnormal behavior registered once the packets reached more than $6$~k packets for different protocols. This analysis is done for different types of protocols and different devices' statuses.

This experiment shows high efficiency in detecting energy consumption attacks as it is not expensive to implement in a smart home device and considers the smart device's resource constraint. Compared to calculating the energy consumption of the devices for detecting energy consumption attacks in smart homes. 
\section{Conclusion and Future work}
\label{ConFU}
The Internet of Things is an internet of smart objects where smart objects communicate with each other. IoT objects are deployed in an open medium with dynamic topology. Due to a lack of infrastructure and centralized management, IoT presents serious vulnerabilities to security attacks, such as energy consumption attacks, as smart devices suffer from resource-constraint. Therefore, security is an essential prerequisite for the real-world deployment of IoT. In this work, we propose a new technique for detecting energy consumption attacks in smart home devices based on the IoT devices' packet rate analysis. This technique considers the received packets related to the IoT devices for different protocols such as TCP, UDP, and the subscribed packets of the MQTT protocol. Therefore with the aim of energy consumption attack detection, the packet reception rate of the IoT devices is calculated and analyzed for each protocol separately or all the protocols simultaneously. In the algorithm, we consider different protocols and different device statuses. Our algorithm shows high efficiency in detecting energy consumption attacks in smart home devices compared to other algorithms that use the current energy consumption measurement for detecting this attack. As this algorithm is easy to use and not expensive to implement, it also considers the resource constraints of smart devices. 

The key observations made from this work present a thorough understanding of the packet reception rate of IoT devices within a home wireless environment. And how the energy consumption attacks could be detected depending on measuring the packet rate received by the smart devices. In the future, we will try to detect the main sources that cause high energy consumption in smart home environments by detecting the attack type.

\bibliographystyle{splncs04}
\bibliography{bibliography}

\end{document}